\documentclass[prb,showpacs,amsmath,amssymb,twocolumn]{revtex4-1}
\usepackage{graphicx,epsfig}

\begin{document}

\title{Non-collinear magnetic ordering in compressed FePd$_3$ ordered alloy: a first principles study}
\author{Y. O. Kvashnin$^{1}$, S. Khmelevskyi$^{2}$, J. Kudrnovsk\'y$^{3}$, A. N. Yaresko$^{4}$, L. Genovese$^{1,5}$, P. Bruno$^{1}$}
\affiliation{$^1$ European Synchrotron Radiation Facility, 6 Rue Jules Horowitz, BP220, 38043 Grenoble Cedex, France}
\affiliation{$^2$ Vienna University of Technology, CMS, Institute for Applied Physics, AT-1020 Vienna, Austria}
\affiliation{$^3$ Institute of Physics, Academy of Sciences of the Czech Republic, Na Slovance 2, CZ-182 21 Prague, Czech Republic}
\affiliation{$^4$ Max-Planck-Institut fur Festkorperforschung, Heisenbergstrasse 1, D-70569 Stuttgart, Germany}
\affiliation{$^5$ Laboratoire de Simulation Atomistique, SP2M, UMR-E CEA / UJF-Grenoble 1, INAC, Grenoble, F-38054, France}

\begin{abstract}
By means of $\textit{ab initio}$ calculations based on the density functional theory we investigated the magnetic phase 
diagram of ordered FePd$_3$ alloy as a function of external pressure. Considering several magnetic configurations 
we concluded that the system under pressure has a tendency toward noncollinear spin alignment. Analysis 
of the Heisenberg exchange parameters Jij revealed strong dependence of iron-iron magnetic couplings on 
polarization of Pd atoms. To take into account that effect we built an extended Heisenberg model with higher 
order (biquadratic) terms. Minimizing the energy of this Hamiltonian, fully parametrized using the results of 
$\textit{ab initio}$ calculations, we found a candidate for a ground state of compressed FePd3 , which can be seen as two 
interpenetrating “triple-Q” phases.
\end{abstract}

\maketitle

\section{Introduction}

Magnetism of Fe alloys is a long standing problem in solid state physics having fundamental importance for modern
technology.\cite{pepperhoff-iron-alloys} 
One of the big challenges is understanding magnetism of iron on a face-centered cubic (fcc) lattice,
which is a key issue toward comprehension of a variety of phenomena, such as spin glass behavior, the Invar anomaly, and a quantitative description of structural transition in steels.

In pure $\gamma$-Fe first-principles calculations predict the
stabilization of a spin spiral state.\cite{Sandratskii-fcc-fe} More generally it has been shown that, for a number of disordered fcc Fe-based binary alloys, such as Fe-Ni and Fe-Pt, the sign of the effective magnetic interaction (ferro- or antiferromagnetic) is dependent on the volume, being ferromagnetic at larger volumes and antiferromagnetic at lower ones.\cite{Olovsson-j0-var,Ruban-FeNi-frust,SK-FePt}
The antiferromagnetic (AFM) interactions on frustrated and chemically disordered lattices lead to the appearance of non-collinear
ground states (GSs) at certain region of volumes, as predicted in work by van Schilfgaarde $\textit{et al.}$ \cite{abrikosov-feni}
Indeed it was shown theoretically that this volume region can be reached under applied pressure.\cite{Abrikosov-Ruban-FeNi,FeNi-magnetovol}

Thanks to the use of diamond anvil cells at synchrotron facilities it has become possible to measure x-ray spectra and investigate properties of matter under extremely high pressure. In today's experiments the values of applied pressure exceed 100 GPa range, giving an opportunity to study ordinary compounds in unconventional conditions. It is of particular interest to investigate Fe-based transition metals alloys under pressure due to the above mentioned volume-dependent magnetic peculiarities of their behaviour. 
A couple of experiments at ultrahigh pressures have been performed on disordered Invar Fe-Ni alloys \cite{Dubrovinsky-FeNi,Matsushita1,Matsushita2,Decremps-FeNi} as well as on Fe$_3$C cementite \cite{sakura-Fe3C}. 
In particulary, the results of these experiments suggest the stabilization of a spin-glass state under pressure in
Fe-Ni and Fe-Pt alloys \cite{Matsushita1,Matsushita2} and an abrupt change of the
magnetic state under some pressure (magnetovolume instability), which is in
general consistent with the prediction of a sharp variation of exchange
couplings with pressure in these systems.\cite{Ruban-FeNi-frust,Olovsson-j0-var,SK-Mohn-invar-nofrust}

FePd$_{3}$ alloy has recently attracted a strong interest due to its Invar
behavior. It was observed experimentally that under an applied pressure of 7
GPa the system shows an anomalously small thermal expansion.\cite{winterrosePRL}
Hence the considered compound demonstrates a pressure-induced Invar effect, reminiscent that of Fe$_{50}$Ni$_{50}$ disordered alloy.\cite{Dubrovinsky-FeNi} 
Moreover, as it will be shown below, non-collinear spin states are stabilized in FePd$_3$ at low volumes, similarly to the Fe$_{50}$Ni$_{50}$ case.

Winterrose $\textit{et al.}$ \cite{winterrosePRL} investigated FePd$_3$ under pressure from both
experimental and theoretical viewpoints. Results of the X-ray diffraction
measurements implied that under the pressure of 12 GPa the system undergoes
a significant volume collapse while preserving its Cu$_3$Au (L1$_2$)
structure. Moreover, under the same applied pressure they observed a
disappearance of quantum beats in a nuclear forward scattering (NFS)
experiment, which indicates the loss of long-range magnetic order in the
system. In order to interpret the results, authors performed a set of
supercell calculations, based on Density Functional Theory (DFT), for few possible magnetic configurations. By
comparing elastic properties of different states with experimental data,
they came to the conclusion that there is a high-spin (HS) to low-spin (LS)
transition taking place under pressure. This hypothesis is supported by the
fact that the obtained magnetic moments in the LS state are of the order of 10$^{-2}$
$\mu_B$ and this can explain the loss of signal in NFS experiment. On the other
hand, the LS phase never had lower total energy than the ferromagnetic (FM)
state in a considered volume range. However, the possibility of paramagnetic
case, i.e. the Curie point under pressure being below room temperature, cannot
be simply excluded.

FePd$_{3}$ is a complex magnetic system, because it is composed of rather 
localized (Fe) and itinerant (Pd) magnetic moments. 
One of the first attempts to account for coexisting magnetism of different characters was
carried out by Mohn and Schwarz.\cite{Mohn-Schwarz} They proposed a model
where local spins produced an effective Weiss field acting on itinerant
magnetic sublattice. The parametrization of the model was based on the results
of $\textit{ab initio}$ calculations. The developed model was applied for Pd-rich Fe$_{x}$Pd$_{1-x}$ ($x<0.1$) alloys and estimated Curie temperatures ($T_{c}$) were found to be in good agreement with experiment.

The problem of induced moments and their influence on magnetic properties of
some Fe-containing alloys was studied in a series of publications by Mryasov $\textit{et al.}$ \cite{mryasov-fept,mryasov-ferh}
For example, it was pointed out that polarized Pt atoms are responsible for
anomalous temperature behavior of magnetocrystalline anisotropy in FePt. 
And the induced magnetization on Rh atoms was shown to play a crucial
role in the phase stability of FeRh.

An extensive first-principles study of iron-palladium compounds was made by
Burzo $\textit{et al.}$\cite{burzo} 
There authors performed calculations using scalar relativistic tight-binding linear muffin-tin orbital method (TB-LMTO).\cite{tb-lmto} The reported values of equilibrium lattice constant and
magnetic moments were in good agreement with available experimental data.
A rough estimation of $T_{c}$ for different Pd concentrations using mean field
approximation (MFA) was not successful, yielding a too low value (while
MFA should normally overestimate the result).

A more accurate approach for calculating $T_c$ in Fe-Pd solutions and
compounds was proposed in Ref.~\onlinecite{polesya}. Pd magnetization was
considered to be proportional to the vector sum of neighboring magnetic
moments. The suggested computational scheme was based on the extended Heisenberg
Hamiltonian with exchange parameters extracted from self-consistent $\textit{ab initio}$ calculation.

Recently another group reported a theoretical study of the magnetism of FePd$_3$ under applied pressure.\cite{dutta-fepd3}
The authors explored various chemically and magnetically disordered states and came to the conclusion that these states can not be the candidates for a GS of the system. 
On the other hand they observed a strong competition between commensurate (FM,AFM) and incommensurate [spin spiral (SS)] magnetic configurations. 
The authors suggest that this is an indication that the system might undergo a transition to some noncollinear state upon the application of pressure.

Mainly motivated by the results of Winterrose $\textit{et al.}$,\cite{winterrosePRL} in the
present work we will show a possible GS of ordered FePd$_{3}$
under pressure and discuss the origins of the magnetic transition.

\section{Methods}

In order to have a realistic description of electronic structure of studied system we carried out a set of self-consistent DFT calculations using PY-LMTO code.\cite{py-lmto}
Correlation effects were treated within Local Spin Density Approximation (LSDA) with the parametrization of Vosko, Wilk and Nusair.\cite{vwn}

The crystalline structure of the compound under consideration is depicted in Fig. 1.
\begin{figure}[!t]
\begin{center}
\includegraphics[angle=0,width=50mm]{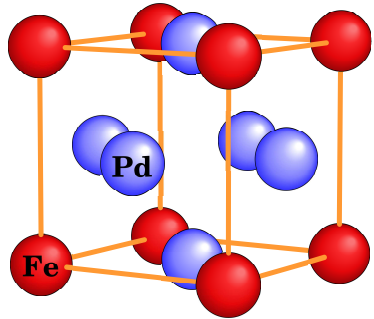}
\caption{Schematic representation of ordered FePd$_3$. 
The alloy has fcc-based L1$_2$ structure, where Fe atoms are located at the corners of the cube and Pd atoms are at the centers of the faces.}
\end{center}
\end{figure}
In the present study we were interested in finding a magnetic GS of a compressed FePd$_3$. 
It is certainly impossible to explore all magnetic configurations as there are an infinite number of them. 
Thus we have investigated the most plausible candidates for the GS by comparing their total energies and examining their stability with respect to infinitesimal deviations of magnetic moments orientations, keeping in mind that this is a probe of the local stability and not the global one.

First we considered possible SS states.
These simulations were carried out on the basis of the generalized Bloch theorem,\cite{Gen-Bloch-theorem} allowing to avoid laborious supercell calculations.
Note, that these calculations are scalar relativistic and therefore there is no coupling between spin degrees of freedom and the crystal lattice. Due to the presence of a global spin rotational invariance, the direction of quantization axis can be chosen arbitrarily. 
We will align it along the $z$ axis.
Hence SS states are defined by four parameters: the propagation vector [$\textbf{Q}=(q_x,q_y,q_z)$] and the angle $\Theta$ formed by magnetization and the $z$ axis. 
Once these parameters are chosen, the magnetization of the iron atom in neighbouring cells is rotated by an angle 
$\phi_i=\textbf{Q}\cdot \textbf{R}$, where $\textbf{R}$ denotes a translational vector. 
Initial phases of Pd moments located at positions $\textbf{t}_{i}$ were set to $\phi^{0}_{i}$=$\textbf{q}$ $\cdot$ $\textbf{t}_{i}$, but were allowed to choose preferred orientation during the self-consistent loop.
The SS calculations are shown in Sec. IV.

Consideration of more complex magnetic phases required construction of appropriate supercells.
In order to accommodate some spin structures, we had to use cells containing up to eight formula units.
The directions of iron magnetic moments were prescribed and frozen during these calculations.
This was done so because the differences in energies associated to spin deviations are rather small ($\sim$meV), so the total energy profile is very shallow in this direction and in addition possesses plenty of local minima.
Due to induced character of Pd magnetic moments, their directions and magnitudes were obtained fully self-consistently.
The corresponding calculations are described in sections V and VI.

The high-temperature paramagnetic (PM) state was modeled by a disordered local moment (DLM)\cite{dlm} configuration.
The effect of magnetic disorder was taken into account by the coherent potential approximation (CPA), as implemented in TB-LMTO-CPA.\cite{turek-gfcpa}
The DLM approach is used to describe properties of the system consisting of randomly distributed magnetic impurities embedded in a non-magnetic medium. Thus induced magnetization on a non-magnetic ions, which is a result of an overlap with spin-polarized bands originating from magnetic atoms, collapses to zero in this phase. 
Therefore, such an approach can be a useful tool for understanding the nature of magnetic moments.

In order to get a deeper insight into magnetic properties, we mapped our system on a classical Heisenberg model of the following form:

\begin{eqnarray}
\label{HH}
\hat H_{exch} = - \sum_{i \neq j} J_{ij} \textbf{e}_i \cdot \textbf{e}_j 
\end{eqnarray}
where $J_{ij}$ denotes the exchange integral between magnetic atoms at sites $i$ and $j$, and $\textbf{e}_i$ and $\textbf{e}_j$ are unit vectors in the directions of the local magnetization on sites $i$ and $j$, respectively.
Exchange parameters were computed using the approach of Lichtenstein $\textit{et al.}$  based on magnetic force theorem:\cite{lichtenstein-exch}
\begin{eqnarray}
\label{licht}
 J_{ij} = \frac{1}{4\pi}  \Im \int^{E_f}_{-\infty}  Tr (\Delta_i G^{\uparrow}_{ij} \Delta_j G^{\downarrow}_{ji})  d\varepsilon
\end{eqnarray}
where $\Delta_i$ is an exchange potential on the $i$-th site and $G^{\sigma}_{ij}$ is an intersite Green's function, which describes propagation of an electron with spin $\sigma=\{\uparrow,\downarrow\}$ from site $i$ to $j$.
Within this method, exchange integrals between two sites are calculated as a response on infinitesimally small deviations of corresponding magnetic moments away from the reference state.
So the assumed magnetic order matters and, as it will be shown later, the effective $J_{ij}$ parameters can be different for various states. The consequences of this will be discussed in more details in Sec. III.
In order to probe the stability of a certain magnetic state, extracted $J$'s were used to find low-energy magnetic excitation (i.e. frozen magnon) spectra, which are the eigenmodes of the considered Hamiltonian.
More detailed description of computational aspects can be found elsewhere.\cite{magnons-jij}

It has to be mentioned that equilibrium volume value and corresponding pressures predicted by LSDA are different from the experimental ones.
In most of the cases LSDA leads to underestimation of the bond length by a few percent.\cite{blaha-lattice-constants}
To avoid any ambiguity, we will work with the experimental lattice parameter for FePd$_{3}$, which is 3.849~\AA~at ambient conditions,\cite{fepd3-a0,winterrosePRL} and its corresponding volume V$_0$ is used throughout the paper.

\section{Heisenberg exchange interactions}

Self-consistent calculations for the DLM state were carried out in TB-LMTO-CPA.
It is found that Fe atoms keep the values of their moments in DLM(PM) state almost unchanged as compared to the magnetically ordered states, thus suggesting a high degree of localization of the iron magnetic moment.
The difference in absolute values of M$_{Fe}$ in DLM and FM states at the equilibrium volume was approximately 2$\%$.
The same correspondence holds under applied pressure and thus there is no tendency toward a drop of the magnetic moment as suggested by the HS-LS scenario.\cite{winterrosePRL} 
Similar results were already reported in a previous study.\cite{dutta-fepd3}

Using obtained TB parameters, we have calculated pair exchange integrals starting from the DLM state.
In this case Fe moments do not have any prescribed orientation and extracted $J-$parameters should reflect the properties of the system in the high-temperature phase above $T_{c}$.
This approach was successfully applied in previous studies.\cite{exch-from-dlm,kudrnov-fe-ir001}

It should be mentioned, however, that application of the magnetic force theorem to systems with induced local moments has limited validity, as was shown by Sandratskii $\textit{et al.}$ \cite{Sandratskii-ind-moms}
In the DLM state small induced Pd moments are reduced to zero due to random orientations of Fe spins. 
We therefore investigate a net effect of iron moments on possible magnetic ground states which may exist at low temperatures.

\begin{figure}[!t]
\includegraphics[angle=0,width=70mm]{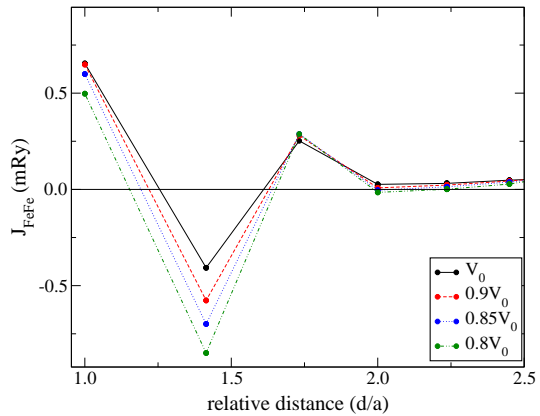}
\caption{The Fe-Fe exchange interactions in FePd$_3$ as functions of the relative inter-atomic distance $d/a$, where $a$ is the lattice parameter.
The results obtained from the DLM reference state calculated for different cell compressions.}
\label{fig:j-dlm}
\end{figure}

In Fig. \ref{fig:j-dlm} we show calculated effective $J_{FeFe}$ parameters as a function of inter-atomic distance for a few fractions of equilibrium volume. The results suggest that the first and second nearest-neighbor (NN) interactions are dominant. While the first NN interactions (6 neighbors) are FM, the second NN (12 neighbors) couplings are AFM. 
Third NN exchange parameters, which are FM, are also important.
Such oscillatory behaviour is due to the Ruderman-Kittel-Kasuya-Yosida (RKKY)\cite{rkky} nature of exchange interactions in metals and is similar to that reported for bcc Fe.\cite{magnons-jij}

We report a strong increase of second NN AFM interactions with pressure while all other couplings depend on volume much more weakly. Moreover, the corresponding neighborhood forms the fcc lattice, which is frustrated for this sign of interaction. 
Here we found that the frustration, being a natural source of noncollinearity in spin systems, effectively increases its contribution at lower volumes. 
We will refer to this fact in the next section.

\begin{figure}[!t]
\includegraphics[angle=0,width=70mm]{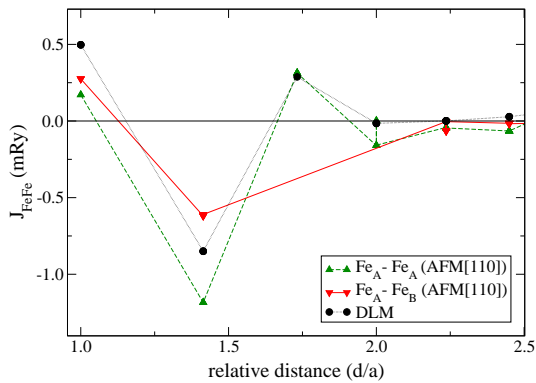}
\caption{Iron-iron exchange parameters obtained from DLM and AFM[110] states for 0.8V$_0$.
Two iron sublattices which appear in the AFM[110] are denoted as Fe$_{A}$ and Fe$_{B}$.}
\label{fig:exch}
\end{figure}

Next we have examined the stability of several magnetic states in a compressed FePd$_3$. 
In Fig. \ref{fig:exch} we show the calculated iron-iron $J_{FeFe}$ exchange parameters extracted from DLM and AFM[110] states.
Using obtained parameters, we computed frozen magnon spectra for each of these states. 
However, all phases demonstrated local magnetic instabilities, indicating that any small perturbation would destroy the state. 
In other words all these magnetic configurations are not even local minima on a phase diagram of the system under consideration.
Nonetheless, we obtained important information: the Fe spins have a tendency toward noncollinearity. 
Motivated by this fact, we examined the energies of spin spiral and other noncollinear states.

\section{Spin spiral calculation}

\begin{figure}[!h]
\begin{center}
\includegraphics[angle=0,width=75mm]{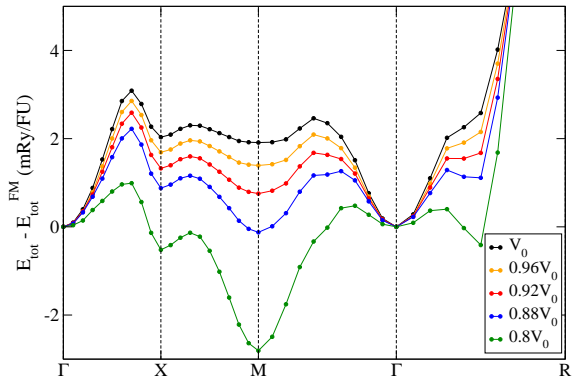}
\caption{Total energies of spin spiral states relative to the energy of the FM state at a given volume in FePd$_3$.
$R-X$ and $R-M$ directions are not shown, as wave-vector dependence of the total energy was found to be monotonic along these paths. The energies are given per chemical formula unit (FU).}
\label{fig:spirals}
\end{center}
\end{figure}

First we explored the manifold of the states characterized by $\Theta$ = 90$^{o}$. 
The energies of these states as a function of a wave vector are shown in Fig. \ref{fig:spirals}.
The results imply that for ambient pressure $\Gamma$ point has the lowest energy among all considered configurations. This is a manifestation of the stability of the FM state.

Nevertheless, there are two local minima at $X$ and $M$ high symmetry points. 
These points correspond to antiferromagnetic states with ordering vectors [100] and [110], respectively. 
One additional minimum is lying along $\Gamma-R$ direction and corresponds to wave vector $\textbf{q}$=$(\frac{\pi}{2a},\frac{\pi}{2a},\frac{\pi}{2a})$.
Such peculiarities of the total energy were already emphasized by another research group.\cite{dutta-fepd3} The overall shape of the energy profile is in fair agreement with those calculations, even though different computational schemes were used.

It is seen that with increasing pressure the stability of the FM state is reduced and at the volume about 0.88V$_0$ we observe the magnetic transition at the point $M$ corresponding to the AFM[110] phase, as was previously reported by Winterrose $\textit{et al.}$\cite{winterrosePRL}
Further compression leads to further destabilization of the ferromagnetic configuration.

It should be noted that iron has a quite rigid magnetic moment in the entire considered volume range: For majority of configurations, as volume is decreased by 20$\%$, its magnetization $M_{Fe}$ lowers by not more than 11$\%$.

Another remarkable fact is that, for a fixed volume, the magnitude of the iron magnetic moment has very similar values among different SS states.
The highest difference was observed between the FM and AFM[110] states and was estimated to be $\approx$0.1 $\mu_B$ per atom.
Meanwhile the value of $M_{Pd}$ strongly depends on its environment as was already pointed out by another group.\cite{polesya}
For example, at V$_0$ volume in FM state all Pd ions possess the magnetic moment of $\approx$0.35 $\mu_B$ per atom.
In the layered AFM[100] state Pd atoms which belong to the same layer as Fe atoms have a magnetization of 0.14 $\mu_B$ per atom, pointing parallel to the iron moment. 
The remaining palladium atoms are non-magnetic.
In the AFM[110] all Pd moments collapse to zero, because each of them is surrounded by an equal number of Fe moments pointing ``up'' ($Fe_{A}$) and ``down'' ($Fe_{B}$).

%\vspace{5 mm}
\begin{figure}[!t]
\begin{center}
\includegraphics[angle=0,width=80mm]{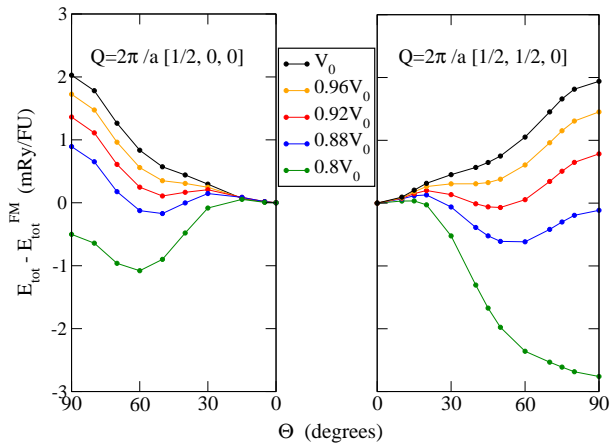}
\caption{Energies of helical spin configurations as a function of the angle between magnetization and the $z$ axis in FePd$_3$. Values are relative to the energy of the FM state at a given volume.}
\label{fig:helix}
\end{center}
\end{figure}
Our next step was an investigation of various helical structures. In this set of calculations we have chosen two $\textbf{Q}$-vectors, corresponding to the lowest states, observed so far, namely AFM[100] and AFM[110]. 
Freezing these three parameters of the SS, we tried to vary the value of the $\Theta$ angle. 
Energies of such magnetic structures are shown in Fig. \ref{fig:helix}.
The results again suggest a reduction of the FM stability with increasing pressure, but for both studied directions ($\textbf{Q}$), we observed a wide range of volumes where helical states are in favor. 
As one compresses the cell up to 0.92V$_0$ the FM phase becomes almost degenerate with two more states, which in the coordinate system ($\textbf{Q},\Theta$) correspond to the points ($X$,50$^{o}$) and ($M$,50$^{o}$). Further volume decrease leads to the destabilization of the FM solution and the states belonging to the family ($M$,$\alpha$) possess the lowest energy in the entire volume range.

At 0.8$V_0$ the AFM[110] phase possesses the lowest energy among spin-polarized states. 
However, we have already pointed out that the calculated frozen magnon dispersion indicated a local instability of this state, i.e. appearance of imaginary eigenvalues in the corresponding spectrum. 
Therefore we deduce that this is not a GS and this fact forced us to investigate more complex states.
Moreover, so far we have considered only coplanar spin structures.

\section{Canted Spin States}

\begin{figure}[!t]
\begin{center}
\includegraphics[angle=0,width=85mm]{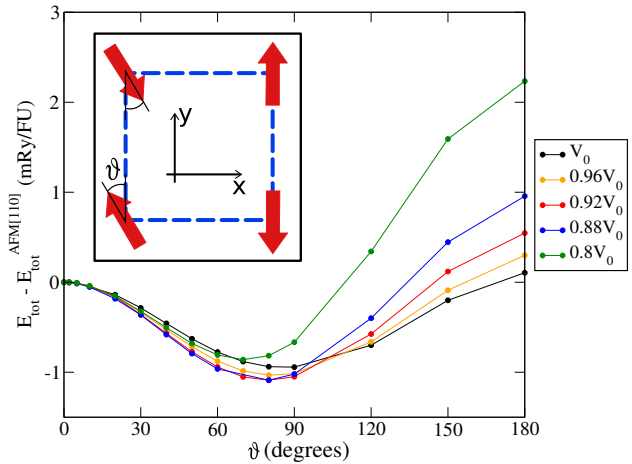}
\caption{Results of the total energy DFT calculations of canted states in ordered FePd$_3$. 
The inset shows the magnetic structure under consideration. Red arrows indicate iron magnetic moments. 
Pd magnetization is not shown.}
\label{fig:canted}
\end{center}
\end{figure}

In order to explore more complex magnetic states, we have constructed a 2$\times$2$\times$1 supercell of FePd$_3$ with four uneqivalent iron sublattices. 
Here we introduce an angle $\theta$ for iron spins, defined in a way shown in the inset of Figure \ref{fig:canted}.
Thus $\theta$ = 0 corresponds to the AFM[110] phase and $\theta$ = 180$^{o}$ - to AFM[100]. 
By tuning $\theta$ one can go continously from one state to another. 
At each volume the calculations were carried out for the angle $\theta$ fixed to a given value. 
A fully unconstrained determination of $\theta$ would be an extremely difficult task, due to the reasons mentioned in Sec. II.

According to the obtained results, corresponding function E$_{tot}$($\theta$) in addition to the points $\theta = \{0,180^{o}\}$ has one more minimum at certain angle ($\theta_0$).
Such angular dependence can not be accounted within classical Heisenberg picture, which should give a ``$\cos{(\theta)}$'' curvature.
Note, that the present shape of the total energy profile was found even without inclusion of spin-orbit coupling (SOC). 
Thus relativistic effects, such as magnetic anisotropy, are not responsible for such behavior.
The situation is reminiscent of the study of iron pnictides,\cite{pnictides-biq} where similar constrained calculations were performed. 
For the latter case it was proposed that the proper E$_{tot}$($\theta$) dependence can be obtained by introducing a higher-order exchange terms to the spin model.

Following this idea, we have built an extended Heisenberg-like model, which aims to catch the essential physical properties of the FePd$_3$ and guide us toward finding a true ground state.

\section{Extended Heisenberg Model and Triple-Q structure}

The influence of Pd moments on stability of the FM state is crucial as was demonstrated by Polesya \textit{et al.}\cite{polesya} and therefore it is essential to take them into account.

Palladium is a rather peculiar element: it is known that already bulk Pd is characterized by a high density of states at the Fermi level, so the Stoner criterion is nearly satisfied. 
Therefore Pd is easily polarized by the contact with neighboring magnetic moments.
Certainly it participates in magnetic interactions and as a result effective couplings between two iron sites in FM phase, where Pd gets polarized, and in AFM one, where it is nonmagnetic, are considerably different (Fig. \ref{fig:exch}). 

The underlying physics can be already understood by looking on the expression of the $J$-parameters in Eq.~\eqref{licht}.
Since the magnitude of Fe magnetic moment is almost configuration-independent, one gets that $|\Delta_{Fe}|$ is approximately the same in all states.
Therefore what gives rise to the observed difference in exchange couplings is the intersite Green's function.
In the FM state an electron, going from one iron site to another, propagates through a strongly polarized medium,
 while in the AFM state this polarization is missing.
Since Pd magnetization is large ($\sim$0.3 $\mu_B$ per atom), being a first-order term in $\Delta_{Fe}$, its disappearance gives a significant impact on $G_{ij}$ and eventually on exchange integrals.
As a result, a classical Heisenberg model with pairwise interactions and bilinear exchange only, cannot properly map the dependence of the total energy on the magnetic configurations and must be extended to include higher-order exchange terms.

Such behaviour was already reported for FeRh compound.\cite{mryasov-ferh}
It is worth emphasizing that this situation is not generic and can be ascribed to be the feature of these 4$d$ elements.
In order to stress this point, we performed additional calculations for a hypothetical system, where palladium was substituted by copper atoms within same geometry of the unit cell.
Opposite to the previous case, Cu is non-magnetic in both AFM[110] and FM states. 
In this case we found that $J_2$ parameters extracted from both configurations almost coincide with each other.
Thus we do not observe any pronounced deviations from the Heisenberg magnet behaviour.

Hence, in order to take into account such peculiarities of the magnetic interactions, we propose an effective model for iron degrees of freedom, which in addition to Eq.~\eqref{HH} contains higher-order exchange terms, originating from the polarization of palladium atoms:
\begin{eqnarray}
\hat H = \hat H_{exch} - \sum_{i \neq j} J^{\prime}_{ij} (\textbf{e}_i \cdot \textbf{e}_j )^2
\end{eqnarray}
where $J^{\prime}$ is a biquadratic exchange parameter.

Thus we are able to build a Heisenberg-like model and fully parametrize it using the results of DFT calculations. For simplicity we considered interactions with first three coordination spheres only ($J_1,J_2,J_3$) as the remainder are much smaller. 
As was already pointed out, the Pd-originated renormalization of $J$-parameters is the most pronounced for next NNs ($J_2$) and therefore we introduce biquadratic term only for this coupling.
\footnote{$J_2^{\prime}$ manifests itself in the AFM state. It gives rise to the difference in the values of the effective parameters between parallel and antiparallel pairs of Fe spins.}
The values which were extracted from the Fig.\ref{fig:exch} and used for the model are the following: $\{J_1,J_2,J^{\prime}_2,J_3\} \rightarrow \{$0.236,-0.902,-0.282,0.29$\}$ (in mRy). 
It is seen that $J^{\prime}_2$ is of the same order as bilinear exchange.
Such situation is atypical, but not unique: for example, a sizable value of the biquadratic term is necessary to explain properties of another class of Fe-based materials.\cite{mc-biquad-pnictides}
Moreover, the present set of parameters describes well the curvature of E$_{tot}$($\theta$) profile, shown in Fig. 6.
The very existence of the minimum $\theta_0$ as well as its position are in fair agreement with those calculations.
This can be viewed as an indication of plausability of the chosen parameters for our model.

\begin{figure}[!t]
\includegraphics[angle=0,width=50mm]{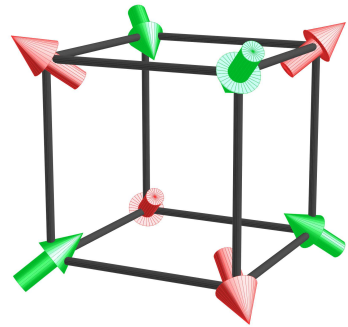}
\caption{Obtained high-pressure magnetic phase of FePd$_3$. Arrows represent iron magnetic moments. 
Four spins of the same color, forming the triple-Q state, point toward the center of the cube, four others - opposite to it. 
Note, that the state is degenerate with respect to simultaneous rotations of all spins through the same angles.}
\label{fig:3q}
\end{figure}

Minimizing the energy of the Hamiltonian (Eq.~(3)) on a 2$\times$2$\times$2 supercell we obtained a new ground state of the system, which is depicted in Fig. \ref{fig:3q}.
The magnetic structure of this state can be seen as two interpenetrating fcc subsystems, whose spins form so-called “triple-Q“ (3Q) states.\cite{Endoh-3Q} 
The angle between each pair of spins is 109$^{o}$28$^{\prime}$, hence the vectors point toward the vertices of an ideal tetrahedron. Each spin out of eight has an identical pair, so its vertices are doubly degenerate.
Analysis of the spin arrangement revealed that $J_3$ coupling (FM), which connects two fcc sublattices, is fully satisfied within such geometry. 
The rest of the interactions compete with each other due to presence of frustration in the system. 
Note that negative sign of $J^{\prime}_2$ means that the interaction favors perpendicular spin orientation. 
This term is the driving force lifting the degeneracy associated with the frustration, and results into stabilization of a certain angle between spins.

It was shown before for fcc-based alloys that stability of triple-Q state with respect to single-Q ones can be related to nonlinear spin interactions in the system\cite{Jo-higher-order} and/or the presence of paramagentic impurities\cite{Cu3Au-3Q-stability}. 
In fact, both of these ingredients are present in FePd$_3$, and hence our findings are consistent with the established physical picture. 
However, to the best of our knowledge the presented spin structure was not observed in alloys with such a low concentration of magnetic ions.

\begin{figure}[!t]
\includegraphics[angle=0,width=75mm]{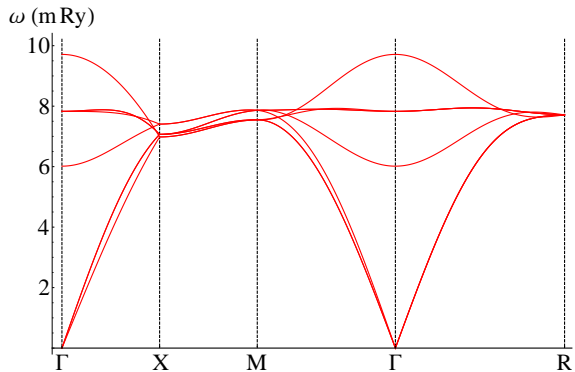}
\caption{Frozen magnon energies of the proposed high-pressure magnetic phase of FePd$_3$.}
\label{fig:magnons}
\end{figure}

In order to check the stability of the obtained configuration, we have computed the corresponding dispersion of low-energy magnetic excitations (Fig.\ref{fig:magnons}).
First of all we confirm that all excitations have a positive energy, meaning that the state is stable. It is worth emphasizing once again, that all states which were studied before showed instabilities in their spectra. Second, one can see few almost degenerate Goldstone modes obeying linear dispersion law at small values of the wave vector $\textbf{q}$. 
We will refer to this fact in the next section.

Finally, we carried out total energy DFT calculation of the 3Q state for different fractions of equilibrium volume. It was confirmed that this noncollinear state becomes lower in energy than FM one at a relative compression rate of 0.96, which is in excellent agreement with the experiment.\cite{winterrosePRL}
The triple-Q state possesses the lowest energy among all studied states in the low-volume region. 
\footnote{We also verified that the inclusion of SOC does not modify the present phase diagram}

Hence we see that the model with the present choice of $J-$parameters describes well the properties of the system next to the GS. It has to be mentioned, however, that this set is unable to reproduce the relative energies of spin spirals depicted in Fig.~\ref{fig:spirals}. 
One of possible explanations is that incommensurate spin states possess a significant Pd magnetization which has to be taken into account explicitly for a proper energy estimation.

\section{Observation of the Triple-Q state}
3Q states are rather difficult to observe experimentally, because their behavior is similar to collinear antiferromagnets. \cite{3Q-vs-AFM}
The measurement which helped to distinguish these two phases was proposed by Kawarazaki $\textit{et al.}$ \cite{3Q-how-to-measure} 
The method, however, requires certain elements which are the sources of the $\gamma$-rays. 
Thus for the case of FePd$_3$ a more useful way would be to use the M\"ossbauer effect in Fe, but this technique can not provide an unequivocal answer if the state is more complex than single-Q type.\cite{moessbauer-on-3Q}

As was already pointed out, first of all we suspect that there should be an abrupt change in the shape of the spin-wave dispersion from parabolic to linear in FePd$_3$ under applied pressure. 
This would be the first indication of existence of the triple-Q state, which is necessary, but not sufficient.

Another external parameter which has to be controlled during an experiment is the temperature. 
In Ref.~\onlinecite{winterrosePRL} a complete vanishing of Fe average magnetization was found in the NFS experiment upon compression. 
One of explanations is that the value of $T_c$ is strongly affected by applied pressure and at a given volume it is pushed below room temperature. 
However, a reliable $\textit{ab initio}$ evaluation of $T_c$ of FM FePd$_3$ would be already a difficult task and seems not to be solved in previous studies.\cite{burzo,polesya}
Recent results on estimation of the $T_c$ in MFA confirm its decrease upon the application of pressure \cite{dutta-fepd3}, but it seems to reach the room temperature at lower volumes than observed experimentally.
It should not be excluded that 3Q configuration might have a different ordering temperature as compared with the FM one. 
This is one of explanations of the loss of the NFS signal.

Another issue with such type of systems is disorder. 
Ordered FePd$_3$ samples are obtained by annealing and subsequent fast cooling (quenching). 
Hence some amount of residual disorder can always appear in a real system. 
It was reported that the value of transition pressure differs from one sample to another.\cite{winterrosePRB} 
This fact may suggest that a small antisite disorder is present and can affect a phase diagram of the compound under consideration. 
In principle, this can give rise to more complicated spin structures. 

\section{Conclusions}
We have studied theoretically the ground-state magnetic properties of FePd$_3$ ordered alloy under the external pressure using  first-principles methods. 
It was deduced that the compound undergoes a magnetic transition from the FM to the Triple-Q state at 0.96 fraction of equilibrium volume. 
Fe atoms possess a significant magnetic moment in the high-pressure phase, unlike the HS-LS scenario suggested earlier. The disappearance of the quantum beats in the NFS experiment\cite{winterrosePRL} may indicate a drop of $T_c$ across the transition.
Essential ingredients which stabilize the 3Q states can be determined: (i) an existence of strong magnetically frustrated couplings in the system, which is $J_2$ in the present case; and (ii) admixture of higher-order interactions, favoring perpendicular spin alignment. 

\section{Acknowledgement}

J.K. acknowledges financial support from the Czech Science Foundations (P204/11/1228) and S.K. acknowledges support from the Austrian Science Fund (FWF) (SFB ViCoM F4109-N13). 
Y.O.K. thanks the Institute of Physics (Prague) for hospitality during his stay.


\begin{thebibliography}{46}%
\makeatletter
\providecommand \@ifxundefined [1]{%
 \@ifx{#1\undefined}
}%
\providecommand \@ifnum [1]{%
 \ifnum #1\expandafter \@firstoftwo
 \else \expandafter \@secondoftwo
 \fi
}%
\providecommand \@ifx [1]{%
 \ifx #1\expandafter \@firstoftwo
 \else \expandafter \@secondoftwo
 \fi
}%
\providecommand \natexlab [1]{#1}%
\providecommand \enquote  [1]{``#1''}%
\providecommand \bibnamefont  [1]{#1}%
\providecommand \bibfnamefont [1]{#1}%
\providecommand \citenamefont [1]{#1}%
\providecommand \href@noop [0]{\@secondoftwo}%
\providecommand \href [0]{\begingroup \@sanitize@url \@href}%
\providecommand \@href[1]{\@@startlink{#1}\@@href}%
\providecommand \@@href[1]{\endgroup#1\@@endlink}%
\providecommand \@sanitize@url [0]{\catcode `\\12\catcode `\$12\catcode
  `\&12\catcode `\#12\catcode `\^12\catcode `\_12\catcode `\%12\relax}%
\providecommand \@@startlink[1]{}%
\providecommand \@@endlink[0]{}%
\providecommand \url  [0]{\begingroup\@sanitize@url \@url }%
\providecommand \@url [1]{\endgroup\@href {#1}{\urlprefix }}%
\providecommand \urlprefix  [0]{URL }%
\providecommand \Eprint [0]{\href }%
\providecommand \doibase [0]{http://dx.doi.org/}%
\providecommand \selectlanguage [0]{\@gobble}%
\providecommand \bibinfo  [0]{\@secondoftwo}%
\providecommand \bibfield  [0]{\@secondoftwo}%
\providecommand \translation [1]{[#1]}%
\providecommand \BibitemOpen [0]{}%
\providecommand \bibitemStop [0]{}%
\providecommand \bibitemNoStop [0]{.\EOS\space}%
\providecommand \EOS [0]{\spacefactor3000\relax}%
\providecommand \BibitemShut  [1]{\csname bibitem#1\endcsname}%
\let\auto@bib@innerbib\@empty
%</preamble>
\bibitem [{\citenamefont {Pepperhoff}\ and\ \citenamefont
  {Acet}(2001)}]{pepperhoff-iron-alloys}%
  \BibitemOpen
  \bibfield  {author} {\bibinfo {author} {\bibfnamefont {W.}~\bibnamefont
  {Pepperhoff}}\ and\ \bibinfo {author} {\bibfnamefont {M.}~\bibnamefont
  {Acet}},\ }\href@noop {} {\emph {\bibinfo {title} {Constitution and Magnetism
  of Iron and its Alloys}}}\ (\bibinfo  {publisher} {Spring-Verlag, Berlin},\
  \bibinfo {year} {2001})\BibitemShut {NoStop}%
\bibitem [{\citenamefont {Sandratskii}(1998)}]{Sandratskii-fcc-fe}%
  \BibitemOpen
  \bibfield  {author} {\bibinfo {author} {\bibfnamefont {L.~M.}\ \bibnamefont
  {Sandratskii}},\ }\href {\doibase 10.1080/000187398243573} {\bibfield
  {journal} {\bibinfo  {journal} {Advances in Physics}\ }\textbf {\bibinfo
  {volume} {47}},\ \bibinfo {pages} {91} (\bibinfo {year} {1998})}\BibitemShut
  {NoStop}%
\bibitem [{\citenamefont {Olovsson}\ and\ \citenamefont
  {Abrikosov}(2005)}]{Olovsson-j0-var}%
  \BibitemOpen
  \bibfield  {author} {\bibinfo {author} {\bibfnamefont {W.}~\bibnamefont
  {Olovsson}}\ and\ \bibinfo {author} {\bibfnamefont {I.~A.}\ \bibnamefont
  {Abrikosov}},\ }\href {\doibase 10.1063/1.1855204} {\bibfield  {journal}
  {\bibinfo  {journal} {Journal of Applied Physics}\ }\textbf {\bibinfo
  {volume} {97}},\ \bibinfo {eid} {10A317} (\bibinfo {year}
  {2005})}\BibitemShut {NoStop}%
\bibitem [{\citenamefont {Ruban}\ \emph {et~al.}(2005)\citenamefont {Ruban},
  \citenamefont {Katsnelson}, \citenamefont {Olovsson}, \citenamefont {Simak},\
  and\ \citenamefont {Abrikosov}}]{Ruban-FeNi-frust}%
  \BibitemOpen
  \bibfield  {author} {\bibinfo {author} {\bibfnamefont {A.~V.}\ \bibnamefont
  {Ruban}}, \bibinfo {author} {\bibfnamefont {M.~I.}\ \bibnamefont
  {Katsnelson}}, \bibinfo {author} {\bibfnamefont {W.}~\bibnamefont
  {Olovsson}}, \bibinfo {author} {\bibfnamefont {S.~I.}\ \bibnamefont {Simak}},
  \ and\ \bibinfo {author} {\bibfnamefont {I.~A.}\ \bibnamefont {Abrikosov}},\
  }\href {\doibase 10.1103/PhysRevB.71.054402} {\bibfield  {journal} {\bibinfo
  {journal} {Phys. Rev. B}\ }\textbf {\bibinfo {volume} {71}},\ \bibinfo
  {pages} {054402} (\bibinfo {year} {2005})}\BibitemShut {NoStop}%
\bibitem [{\citenamefont {Khmelevskyi}\ and\ \citenamefont
  {Mohn}(2003)}]{SK-FePt}%
  \BibitemOpen
  \bibfield  {author} {\bibinfo {author} {\bibfnamefont {S.}~\bibnamefont
  {Khmelevskyi}}\ and\ \bibinfo {author} {\bibfnamefont {P.}~\bibnamefont
  {Mohn}},\ }\href {\doibase 10.1103/PhysRevB.68.214412} {\bibfield  {journal}
  {\bibinfo  {journal} {Phys. Rev. B}\ }\textbf {\bibinfo {volume} {68}},\
  \bibinfo {pages} {214412} (\bibinfo {year} {2003})}\BibitemShut {NoStop}%
\bibitem [{\citenamefont {van Schilfgaarde}\ \emph {et~al.}(1999)\citenamefont
  {van Schilfgaarde}, \citenamefont {Abrikosov},\ and\ \citenamefont
  {Johansson}}]{abrikosov-feni}%
  \BibitemOpen
  \bibfield  {author} {\bibinfo {author} {\bibfnamefont {M.}~\bibnamefont {van
  Schilfgaarde}}, \bibinfo {author} {\bibfnamefont {I.}~\bibnamefont
  {Abrikosov}}, \ and\ \bibinfo {author} {\bibfnamefont {B.}~\bibnamefont
  {Johansson}},\ }\href {\doibase 10.1038/21848} {\bibfield  {journal}
  {\bibinfo  {journal} {Nature}\ }\textbf {\bibinfo {volume} {400}},\ \bibinfo
  {pages} {46} (\bibinfo {year} {1999})}\BibitemShut {NoStop}%
\bibitem [{\citenamefont {Abrikosov}\ \emph {et~al.}(2007)\citenamefont
  {Abrikosov}, \citenamefont {Kissavos}, \citenamefont {Liot}, \citenamefont
  {Alling}, \citenamefont {Simak}, \citenamefont {Peil},\ and\ \citenamefont
  {Ruban}}]{Abrikosov-Ruban-FeNi}%
  \BibitemOpen
  \bibfield  {author} {\bibinfo {author} {\bibfnamefont {I.~A.}\ \bibnamefont
  {Abrikosov}}, \bibinfo {author} {\bibfnamefont {A.~E.}\ \bibnamefont
  {Kissavos}}, \bibinfo {author} {\bibfnamefont {F.}~\bibnamefont {Liot}},
  \bibinfo {author} {\bibfnamefont {B.}~\bibnamefont {Alling}}, \bibinfo
  {author} {\bibfnamefont {S.~I.}\ \bibnamefont {Simak}}, \bibinfo {author}
  {\bibfnamefont {O.}~\bibnamefont {Peil}}, \ and\ \bibinfo {author}
  {\bibfnamefont {A.~V.}\ \bibnamefont {Ruban}},\ }\href {\doibase
  10.1103/PhysRevB.76.014434} {\bibfield  {journal} {\bibinfo  {journal} {Phys.
  Rev. B}\ }\textbf {\bibinfo {volume} {76}},\ \bibinfo {pages} {014434}
  (\bibinfo {year} {2007})}\BibitemShut {NoStop}%
\bibitem [{\citenamefont {Ruban}\ \emph {et~al.}(2007)\citenamefont {Ruban},
  \citenamefont {Khmelevskyi}, \citenamefont {Mohn},\ and\ \citenamefont
  {Johansson}}]{FeNi-magnetovol}%
  \BibitemOpen
  \bibfield  {author} {\bibinfo {author} {\bibfnamefont {A.~V.}\ \bibnamefont
  {Ruban}}, \bibinfo {author} {\bibfnamefont {S.}~\bibnamefont {Khmelevskyi}},
  \bibinfo {author} {\bibfnamefont {P.}~\bibnamefont {Mohn}}, \ and\ \bibinfo
  {author} {\bibfnamefont {B.}~\bibnamefont {Johansson}},\ }\href {\doibase
  10.1103/PhysRevB.76.014420} {\bibfield  {journal} {\bibinfo  {journal} {Phys.
  Rev. B}\ }\textbf {\bibinfo {volume} {76}},\ \bibinfo {pages} {014420}
  (\bibinfo {year} {2007})}\BibitemShut {NoStop}%
\bibitem [{\citenamefont {Dubrovinsky}\ \emph {et~al.}(2001)\citenamefont
  {Dubrovinsky}, \citenamefont {Dubrovinskaia}, \citenamefont {Abrikosov},
  \citenamefont {Vennstr{\"o}m}, \citenamefont {Westman}, \citenamefont
  {Carlson}, \citenamefont {van Schilfgaarde},\ and\ \citenamefont
  {Johansson}}]{Dubrovinsky-FeNi}%
  \BibitemOpen
  \bibfield  {author} {\bibinfo {author} {\bibfnamefont {L.}~\bibnamefont
  {Dubrovinsky}}, \bibinfo {author} {\bibfnamefont {N.}~\bibnamefont
  {Dubrovinskaia}}, \bibinfo {author} {\bibfnamefont {I.~A.}\ \bibnamefont
  {Abrikosov}}, \bibinfo {author} {\bibfnamefont {M.}~\bibnamefont
  {Vennstr{\"o}m}}, \bibinfo {author} {\bibfnamefont {F.}~\bibnamefont
  {Westman}}, \bibinfo {author} {\bibfnamefont {S.}~\bibnamefont {Carlson}},
  \bibinfo {author} {\bibfnamefont {M.}~\bibnamefont {van Schilfgaarde}}, \
  and\ \bibinfo {author} {\bibfnamefont {B.}~\bibnamefont {Johansson}},\ }\href
  {\doibase 10.1103/PhysRevLett.86.4851} {\bibfield  {journal} {\bibinfo
  {journal} {Phys. Rev. Lett.}\ }\textbf {\bibinfo {volume} {86}},\ \bibinfo
  {pages} {4851} (\bibinfo {year} {2001})}\BibitemShut {NoStop}%
\bibitem [{\citenamefont {Matsushita}\ \emph
  {et~al.}(2003{\natexlab{a}})\citenamefont {Matsushita}, \citenamefont {Endo},
  \citenamefont {Miura},\ and\ \citenamefont {Ono}}]{Matsushita1}%
  \BibitemOpen
  \bibfield  {author} {\bibinfo {author} {\bibfnamefont {M.}~\bibnamefont
  {Matsushita}}, \bibinfo {author} {\bibfnamefont {S.}~\bibnamefont {Endo}},
  \bibinfo {author} {\bibfnamefont {K.}~\bibnamefont {Miura}}, \ and\ \bibinfo
  {author} {\bibfnamefont {F.}~\bibnamefont {Ono}},\ }\href {\doibase
  10.1016/S0304-8853(03)00287-7} {\bibfield  {journal} {\bibinfo  {journal}
  {Journal of Magnetism and Magnetic Materials}\ }\textbf {\bibinfo {volume}
  {265}},\ \bibinfo {pages} {352 } (\bibinfo {year}
  {2003}{\natexlab{a}})}\BibitemShut {NoStop}%
\bibitem [{\citenamefont {Matsushita}\ \emph
  {et~al.}(2003{\natexlab{b}})\citenamefont {Matsushita}, \citenamefont {Endo},
  \citenamefont {Miura},\ and\ \citenamefont {Ono}}]{Matsushita2}%
  \BibitemOpen
  \bibfield  {author} {\bibinfo {author} {\bibfnamefont {M.}~\bibnamefont
  {Matsushita}}, \bibinfo {author} {\bibfnamefont {S.}~\bibnamefont {Endo}},
  \bibinfo {author} {\bibfnamefont {K.}~\bibnamefont {Miura}}, \ and\ \bibinfo
  {author} {\bibfnamefont {F.}~\bibnamefont {Ono}},\ }\href {\doibase
  10.1016/S0304-8853(02)01342-2} {\bibfield  {journal} {\bibinfo  {journal}
  {Journal of Magnetism and Magnetic Materials}\ }\textbf {\bibinfo {volume}
  {260}},\ \bibinfo {pages} {371 } (\bibinfo {year}
  {2003}{\natexlab{b}})}\BibitemShut {NoStop}%
\bibitem [{\citenamefont {Decremps}\ and\ \citenamefont
  {Nataf}(2004)}]{Decremps-FeNi}%
  \BibitemOpen
  \bibfield  {author} {\bibinfo {author} {\bibfnamefont {F.}~\bibnamefont
  {Decremps}}\ and\ \bibinfo {author} {\bibfnamefont {L.}~\bibnamefont
  {Nataf}},\ }\href {\doibase 10.1103/PhysRevLett.92.157204} {\bibfield
  {journal} {\bibinfo  {journal} {Phys. Rev. Lett.}\ }\textbf {\bibinfo
  {volume} {92}},\ \bibinfo {pages} {157204} (\bibinfo {year}
  {2004})}\BibitemShut {NoStop}%
\bibitem [{\citenamefont {Duman}\ \emph {et~al.}(2005)\citenamefont {Duman},
  \citenamefont {Acet}, \citenamefont {Wassermann}, \citenamefont {Iti\'e},
  \citenamefont {Baudelet}, \citenamefont {Mathon},\ and\ \citenamefont
  {Pascarelli}}]{sakura-Fe3C}%
  \BibitemOpen
  \bibfield  {author} {\bibinfo {author} {\bibfnamefont {E.}~\bibnamefont
  {Duman}}, \bibinfo {author} {\bibfnamefont {M.}~\bibnamefont {Acet}},
  \bibinfo {author} {\bibfnamefont {E.~F.}\ \bibnamefont {Wassermann}},
  \bibinfo {author} {\bibfnamefont {J.~P.}\ \bibnamefont {Iti\'e}}, \bibinfo
  {author} {\bibfnamefont {F.}~\bibnamefont {Baudelet}}, \bibinfo {author}
  {\bibfnamefont {O.}~\bibnamefont {Mathon}}, \ and\ \bibinfo {author}
  {\bibfnamefont {S.}~\bibnamefont {Pascarelli}},\ }\href {\doibase
  10.1103/PhysRevLett.94.075502} {\bibfield  {journal} {\bibinfo  {journal}
  {Phys. Rev. Lett.}\ }\textbf {\bibinfo {volume} {94}},\ \bibinfo {pages}
  {075502} (\bibinfo {year} {2005})}\BibitemShut {NoStop}%
\bibitem [{\citenamefont {Khmelevskyi}\ and\ \citenamefont
  {Mohn}(2010)}]{SK-Mohn-invar-nofrust}%
  \BibitemOpen
  \bibfield  {author} {\bibinfo {author} {\bibfnamefont {S.}~\bibnamefont
  {Khmelevskyi}}\ and\ \bibinfo {author} {\bibfnamefont {P.}~\bibnamefont
  {Mohn}},\ }\href {\doibase 10.1103/PhysRevB.82.134402} {\bibfield  {journal}
  {\bibinfo  {journal} {Phys. Rev. B}\ }\textbf {\bibinfo {volume} {82}},\
  \bibinfo {pages} {134402} (\bibinfo {year} {2010})}\BibitemShut {NoStop}%
\bibitem [{\citenamefont {Winterrose}\ \emph {et~al.}(2009)\citenamefont
  {Winterrose}, \citenamefont {Lucas}, \citenamefont {Yue}, \citenamefont
  {Halevy}, \citenamefont {Mauger}, \citenamefont {Mu\~noz}, \citenamefont
  {Hu}, \citenamefont {Lerche},\ and\ \citenamefont {Fultz}}]{winterrosePRL}%
  \BibitemOpen
  \bibfield  {author} {\bibinfo {author} {\bibfnamefont {M.~L.}\ \bibnamefont
  {Winterrose}}, \bibinfo {author} {\bibfnamefont {M.~S.}\ \bibnamefont
  {Lucas}}, \bibinfo {author} {\bibfnamefont {A.~F.}\ \bibnamefont {Yue}},
  \bibinfo {author} {\bibfnamefont {I.}~\bibnamefont {Halevy}}, \bibinfo
  {author} {\bibfnamefont {L.}~\bibnamefont {Mauger}}, \bibinfo {author}
  {\bibfnamefont {J.~A.}\ \bibnamefont {Mu\~noz}}, \bibinfo {author}
  {\bibfnamefont {J.}~\bibnamefont {Hu}}, \bibinfo {author} {\bibfnamefont
  {M.}~\bibnamefont {Lerche}}, \ and\ \bibinfo {author} {\bibfnamefont
  {B.}~\bibnamefont {Fultz}},\ }\href {\doibase 10.1103/PhysRevLett.102.237202}
  {\bibfield  {journal} {\bibinfo  {journal} {Phys. Rev. Lett.}\ }\textbf
  {\bibinfo {volume} {102}},\ \bibinfo {pages} {237202} (\bibinfo {year}
  {2009})}\BibitemShut {NoStop}%
\bibitem [{\citenamefont {Mohn}\ and\ \citenamefont
  {Schwarz}(1993)}]{Mohn-Schwarz}%
  \BibitemOpen
  \bibfield  {author} {\bibinfo {author} {\bibfnamefont {P.}~\bibnamefont
  {Mohn}}\ and\ \bibinfo {author} {\bibfnamefont {K.}~\bibnamefont {Schwarz}},\
  }\href {http://stacks.iop.org/0953-8984/5/i=29/a=007} {\bibfield  {journal}
  {\bibinfo  {journal} {Journal of Physics: Condensed Matter}\ }\textbf
  {\bibinfo {volume} {5}},\ \bibinfo {pages} {5099} (\bibinfo {year}
  {1993})}\BibitemShut {NoStop}%
\bibitem [{\citenamefont {Mryasov}\ \emph {et~al.}(2005)\citenamefont
  {Mryasov}, \citenamefont {Nowak}, \citenamefont {Guslienko},\ and\
  \citenamefont {Chantrell}}]{mryasov-fept}%
  \BibitemOpen
  \bibfield  {author} {\bibinfo {author} {\bibfnamefont {O.~N.}\ \bibnamefont
  {Mryasov}}, \bibinfo {author} {\bibfnamefont {U.}~\bibnamefont {Nowak}},
  \bibinfo {author} {\bibfnamefont {K.~Y.}\ \bibnamefont {Guslienko}}, \ and\
  \bibinfo {author} {\bibfnamefont {R.~W.}\ \bibnamefont {Chantrell}},\ }\href
  {http://stacks.iop.org/0295-5075/69/i=5/a=805} {\bibfield  {journal}
  {\bibinfo  {journal} {EPL (Europhysics Letters)}\ }\textbf {\bibinfo {volume}
  {69}},\ \bibinfo {pages} {805} (\bibinfo {year} {2005})}\BibitemShut
  {NoStop}%
\bibitem [{\citenamefont {Mryasov}(2005)}]{mryasov-ferh}%
  \BibitemOpen
  \bibfield  {author} {\bibinfo {author} {\bibfnamefont {O.~N.}\ \bibnamefont
  {Mryasov}},\ }\href {\doibase 10.1080/01411590412331316591} {\bibfield
  {journal} {\bibinfo  {journal} {Phase Transitions}\ }\textbf {\bibinfo
  {volume} {78}},\ \bibinfo {pages} {197} (\bibinfo {year} {2005})}\BibitemShut
  {NoStop}%
\bibitem [{\citenamefont {Burzo}\ and\ \citenamefont {Vlaic}(2010)}]{burzo}%
  \BibitemOpen
  \bibfield  {author} {\bibinfo {author} {\bibfnamefont {E.}~\bibnamefont
  {Burzo}}\ and\ \bibinfo {author} {\bibfnamefont {P.}~\bibnamefont {Vlaic}},\
  }\href@noop {} {\bibfield  {journal} {\bibinfo  {journal} {J. Optoelectron.
  Adv. Mater}\ }\textbf {\bibinfo {volume} {12}},\ \bibinfo {pages} {1869}
  (\bibinfo {year} {2010})}\BibitemShut {NoStop}%
\bibitem [{\citenamefont {Andersen}\ and\ \citenamefont
  {Jepsen}(1984)}]{tb-lmto}%
  \BibitemOpen
  \bibfield  {author} {\bibinfo {author} {\bibfnamefont {O.~K.}\ \bibnamefont
  {Andersen}}\ and\ \bibinfo {author} {\bibfnamefont {O.}~\bibnamefont
  {Jepsen}},\ }\href {\doibase 10.1103/PhysRevLett.53.2571} {\bibfield
  {journal} {\bibinfo  {journal} {Phys. Rev. Lett.}\ }\textbf {\bibinfo
  {volume} {53}},\ \bibinfo {pages} {2571} (\bibinfo {year}
  {1984})}\BibitemShut {NoStop}%
\bibitem [{\citenamefont {Polesya}\ \emph {et~al.}(2010)\citenamefont
  {Polesya}, \citenamefont {Mankovsky}, \citenamefont {Sipr}, \citenamefont
  {Meindl}, \citenamefont {Strunk},\ and\ \citenamefont {Ebert}}]{polesya}%
  \BibitemOpen
  \bibfield  {author} {\bibinfo {author} {\bibfnamefont {S.}~\bibnamefont
  {Polesya}}, \bibinfo {author} {\bibfnamefont {S.}~\bibnamefont {Mankovsky}},
  \bibinfo {author} {\bibfnamefont {O.}~\bibnamefont {Sipr}}, \bibinfo {author}
  {\bibfnamefont {W.}~\bibnamefont {Meindl}}, \bibinfo {author} {\bibfnamefont
  {C.}~\bibnamefont {Strunk}}, \ and\ \bibinfo {author} {\bibfnamefont
  {H.}~\bibnamefont {Ebert}},\ }\href {\doibase 10.1103/PhysRevB.82.214409}
  {\bibfield  {journal} {\bibinfo  {journal} {Phys. Rev. B}\ }\textbf {\bibinfo
  {volume} {82}},\ \bibinfo {pages} {214409} (\bibinfo {year}
  {2010})}\BibitemShut {NoStop}%
\bibitem [{\citenamefont {Dutta}\ \emph {et~al.}(2012)\citenamefont {Dutta},
  \citenamefont {Bhandary}, \citenamefont {Ghosh},\ and\ \citenamefont
  {Sanyal}}]{dutta-fepd3}%
  \BibitemOpen
  \bibfield  {author} {\bibinfo {author} {\bibfnamefont {B.}~\bibnamefont
  {Dutta}}, \bibinfo {author} {\bibfnamefont {S.}~\bibnamefont {Bhandary}},
  \bibinfo {author} {\bibfnamefont {S.}~\bibnamefont {Ghosh}}, \ and\ \bibinfo
  {author} {\bibfnamefont {B.}~\bibnamefont {Sanyal}},\ }\href {\doibase
  10.1103/PhysRevB.86.024419} {\bibfield  {journal} {\bibinfo  {journal} {Phys.
  Rev. B}\ }\textbf {\bibinfo {volume} {86}},\ \bibinfo {pages} {024419}
  (\bibinfo {year} {2012})}\BibitemShut {NoStop}%
\bibitem [{\citenamefont {Antonov}\ \emph {et~al.}(2004)\citenamefont
  {Antonov}, \citenamefont {Harmon},\ and\ \citenamefont {Yaresko}}]{py-lmto}%
  \BibitemOpen
  \bibfield  {author} {\bibinfo {author} {\bibfnamefont {V.}~\bibnamefont
  {Antonov}}, \bibinfo {author} {\bibfnamefont {B.}~\bibnamefont {Harmon}}, \
  and\ \bibinfo {author} {\bibfnamefont {A.}~\bibnamefont {Yaresko}},\
  }\href@noop {} {\emph {\bibinfo {title} {Electronic Structure and
  Magneto-Optical Properties of Solids}}}\ (\bibinfo  {publisher} {Kluwer
  Academic,Dordrecht},\ \bibinfo {year} {2004})\BibitemShut {NoStop}%
\bibitem [{\citenamefont {Vosko}\ \emph {et~al.}(1980)\citenamefont {Vosko},
  \citenamefont {Wilk},\ and\ \citenamefont {Nusair}}]{vwn}%
  \BibitemOpen
  \bibfield  {author} {\bibinfo {author} {\bibfnamefont {S.~H.}\ \bibnamefont
  {Vosko}}, \bibinfo {author} {\bibfnamefont {L.}~\bibnamefont {Wilk}}, \ and\
  \bibinfo {author} {\bibfnamefont {M.}~\bibnamefont {Nusair}},\ }\href
  {\doibase 10.1139/p80-159} {\bibfield  {journal} {\bibinfo  {journal}
  {Canadian Journal of Physics}\ }\textbf {\bibinfo {volume} {58}},\ \bibinfo
  {pages} {1200} (\bibinfo {year} {1980})}\BibitemShut {NoStop}%
\bibitem [{\citenamefont {Sandratskii}(1991)}]{Gen-Bloch-theorem}%
  \BibitemOpen
  \bibfield  {author} {\bibinfo {author} {\bibfnamefont {L.~M.}\ \bibnamefont
  {Sandratskii}},\ }\href {\doibase 10.1088/0953-8984/3/44/004} {\bibfield
  {journal} {\bibinfo  {journal} {Journal of Physics: Condensed Matter}\
  }\textbf {\bibinfo {volume} {3}},\ \bibinfo {pages} {8565} (\bibinfo {year}
  {1991})}\BibitemShut {NoStop}%
\bibitem [{\citenamefont {Gyorffy}\ \emph {et~al.}(1985)\citenamefont
  {Gyorffy}, \citenamefont {Pindor}, \citenamefont {Staunton}, \citenamefont
  {Stocks},\ and\ \citenamefont {Winter}}]{dlm}%
  \BibitemOpen
  \bibfield  {author} {\bibinfo {author} {\bibfnamefont {B.}~\bibnamefont
  {Gyorffy}}, \bibinfo {author} {\bibfnamefont {A.}~\bibnamefont {Pindor}},
  \bibinfo {author} {\bibfnamefont {J.}~\bibnamefont {Staunton}}, \bibinfo
  {author} {\bibfnamefont {G.}~\bibnamefont {Stocks}}, \ and\ \bibinfo {author}
  {\bibfnamefont {H.}~\bibnamefont {Winter}},\ }\href {\doibase
  10.1088/0305-4608/15/6/018} {\bibfield  {journal} {\bibinfo  {journal}
  {Journal of Physics F: Metal Physics}\ }\textbf {\bibinfo {volume} {15}},\
  \bibinfo {pages} {1337} (\bibinfo {year} {1985})}\BibitemShut {NoStop}%
\bibitem [{\citenamefont {Turek}\ \emph {et~al.}(1997)\citenamefont {Turek},
  \citenamefont {Drchal}, \citenamefont {Kudrnovsky}, \citenamefont {Sob},\
  and\ \citenamefont {Weinberger}}]{turek-gfcpa}%
  \BibitemOpen
  \bibfield  {author} {\bibinfo {author} {\bibfnamefont {I.}~\bibnamefont
  {Turek}}, \bibinfo {author} {\bibfnamefont {V.}~\bibnamefont {Drchal}},
  \bibinfo {author} {\bibfnamefont {J.}~\bibnamefont {Kudrnovsky}}, \bibinfo
  {author} {\bibfnamefont {M.}~\bibnamefont {Sob}}, \ and\ \bibinfo {author}
  {\bibfnamefont {P.}~\bibnamefont {Weinberger}},\ }\href@noop {} {\bibfield
  {journal} {\bibinfo  {journal} {Electronic Structure of Disordered Alloys,
  Surfaces and Interfaces}\ ,\ \bibinfo {pages} {Kluwer Academic, Boston}}
  (\bibinfo {year} {1997})}\BibitemShut {NoStop}%
\bibitem [{\citenamefont {Liechtenstein}\ \emph {et~al.}(1987)\citenamefont
  {Liechtenstein}, \citenamefont {Katsnelson}, \citenamefont {Antropov},\ and\
  \citenamefont {Gubanov}}]{lichtenstein-exch}%
  \BibitemOpen
  \bibfield  {author} {\bibinfo {author} {\bibfnamefont {A.}~\bibnamefont
  {Liechtenstein}}, \bibinfo {author} {\bibfnamefont {M.}~\bibnamefont
  {Katsnelson}}, \bibinfo {author} {\bibfnamefont {V.}~\bibnamefont
  {Antropov}}, \ and\ \bibinfo {author} {\bibfnamefont {V.}~\bibnamefont
  {Gubanov}},\ }\href {\doibase 10.1016/0304-8853(87)90721-9} {\bibfield
  {journal} {\bibinfo  {journal} {Journal of Magnetism and Magnetic Materials}\
  }\textbf {\bibinfo {volume} {67}},\ \bibinfo {pages} {65 } (\bibinfo {year}
  {1987})}\BibitemShut {NoStop}%
\bibitem [{\citenamefont {Pajda}\ \emph {et~al.}(2001)\citenamefont {Pajda},
  \citenamefont {Kudrnovsk{\'y}}, \citenamefont {Turek}, \citenamefont
  {Drchal},\ and\ \citenamefont {Bruno}}]{magnons-jij}%
  \BibitemOpen
  \bibfield  {author} {\bibinfo {author} {\bibfnamefont {M.}~\bibnamefont
  {Pajda}}, \bibinfo {author} {\bibfnamefont {J.}~\bibnamefont
  {Kudrnovsk{\'y}}}, \bibinfo {author} {\bibfnamefont {I.}~\bibnamefont
  {Turek}}, \bibinfo {author} {\bibfnamefont {V.}~\bibnamefont {Drchal}}, \
  and\ \bibinfo {author} {\bibfnamefont {P.}~\bibnamefont {Bruno}},\ }\href
  {\doibase 10.1103/PhysRevB.64.174402} {\bibfield  {journal} {\bibinfo
  {journal} {Phys. Rev. B}\ }\textbf {\bibinfo {volume} {64}},\ \bibinfo
  {pages} {174402} (\bibinfo {year} {2001})}\BibitemShut {NoStop}%
\bibitem [{\citenamefont {Haas}\ \emph {et~al.}(2009)\citenamefont {Haas},
  \citenamefont {Tran},\ and\ \citenamefont {Blaha}}]{blaha-lattice-constants}%
  \BibitemOpen
  \bibfield  {author} {\bibinfo {author} {\bibfnamefont {P.}~\bibnamefont
  {Haas}}, \bibinfo {author} {\bibfnamefont {F.}~\bibnamefont {Tran}}, \ and\
  \bibinfo {author} {\bibfnamefont {P.}~\bibnamefont {Blaha}},\ }\href
  {\doibase 10.1103/PhysRevB.79.085104} {\bibfield  {journal} {\bibinfo
  {journal} {Phys. Rev. B}\ }\textbf {\bibinfo {volume} {79}},\ \bibinfo
  {pages} {085104} (\bibinfo {year} {2009})}\BibitemShut {NoStop}%
\bibitem [{\citenamefont {Hansen}(1958)}]{fepd3-a0}%
  \BibitemOpen
  \bibfield  {author} {\bibinfo {author} {\bibfnamefont {M.}~\bibnamefont
  {Hansen}},\ }\href@noop {} {\emph {\bibinfo {title} {Constitution of Binary
  Alloys, Metallurgy and Metallurgical Engineering Series}}}\ (\bibinfo
  {publisher} {McGraw-Hill, New York},\ \bibinfo {year} {1958})\BibitemShut
  {NoStop}%
\bibitem [{\citenamefont {Bose}\ \emph {et~al.}(2010)\citenamefont {Bose},
  \citenamefont {Kudrnovsk\'y}, \citenamefont {Drchal},\ and\ \citenamefont
  {Turek}}]{exch-from-dlm}%
  \BibitemOpen
  \bibfield  {author} {\bibinfo {author} {\bibfnamefont {S.~K.}\ \bibnamefont
  {Bose}}, \bibinfo {author} {\bibfnamefont {J.}~\bibnamefont {Kudrnovsk\'y}},
  \bibinfo {author} {\bibfnamefont {V.}~\bibnamefont {Drchal}}, \ and\ \bibinfo
  {author} {\bibfnamefont {I.}~\bibnamefont {Turek}},\ }\href {\doibase
  10.1103/PhysRevB.82.174402} {\bibfield  {journal} {\bibinfo  {journal} {Phys.
  Rev. B}\ }\textbf {\bibinfo {volume} {82}},\ \bibinfo {pages} {174402}
  (\bibinfo {year} {2010})}\BibitemShut {NoStop}%
\bibitem [{\citenamefont {Kudrnovsk\'y}\ \emph {et~al.}(2009)\citenamefont
  {Kudrnovsk\'y}, \citenamefont {M\'aca}, \citenamefont {Turek},\ and\
  \citenamefont {Redinger}}]{kudrnov-fe-ir001}%
  \BibitemOpen
  \bibfield  {author} {\bibinfo {author} {\bibfnamefont {J.}~\bibnamefont
  {Kudrnovsk\'y}}, \bibinfo {author} {\bibfnamefont {F.}\ \bibnamefont
  {M\'aca}}, \bibinfo {author} {\bibfnamefont {I.}~\bibnamefont {Turek}}, \
  and\ \bibinfo {author} {\bibfnamefont {J.}~\bibnamefont {Redinger}},\ }\href
  {\doibase 10.1103/PhysRevB.80.064405} {\bibfield  {journal} {\bibinfo
  {journal} {Phys. Rev. B}\ }\textbf {\bibinfo {volume} {80}},\ \bibinfo
  {pages} {064405} (\bibinfo {year} {2009})}\BibitemShut {NoStop}%
\bibitem [{\citenamefont {Sandratskii}\ \emph {et~al.}(2007)\citenamefont
  {Sandratskii}, \citenamefont {Singer},\ and\ \citenamefont {\ifmmode
  \mbox{\c{S}}\else \c{S}\fi{}a\ifmmode \mbox{\c{s}}\else \c{s}\fi{}\ifmmode
  \imath \else \i \fi{}o\ifmmode~\breve{g}\else
  \u{g}\fi{}lu}}]{Sandratskii-ind-moms}%
  \BibitemOpen
  \bibfield  {author} {\bibinfo {author} {\bibfnamefont {L.~M.}\ \bibnamefont
  {Sandratskii}}, \bibinfo {author} {\bibfnamefont {R.}~\bibnamefont {Singer}},
  \ and\ \bibinfo {author} {\bibfnamefont {E.}~\bibnamefont {\ifmmode
  \mbox{\c{S}}\else \c{S}\fi{}a\ifmmode \mbox{\c{s}}\else \c{s}\fi{}\ifmmode
  \imath \else \i \fi{}o\ifmmode~\breve{g}\else \u{g}\fi{}lu}},\ }\href
  {\doibase 10.1103/PhysRevB.76.184406} {\bibfield  {journal} {\bibinfo
  {journal} {Phys. Rev. B}\ }\textbf {\bibinfo {volume} {76}},\ \bibinfo
  {pages} {184406} (\bibinfo {year} {2007})}\BibitemShut {NoStop}%
\bibitem [{\citenamefont {Ruderman}\ and\ \citenamefont {Kittel}(1954)}]{rkky}%
  \BibitemOpen
  \bibfield  {author} {\bibinfo {author} {\bibfnamefont {M.~A.}\ \bibnamefont
  {Ruderman}}\ and\ \bibinfo {author} {\bibfnamefont {C.}~\bibnamefont
  {Kittel}},\ }\href {\doibase 10.1103/PhysRev.96.99} {\bibfield  {journal}
  {\bibinfo  {journal} {Phys. Rev.}\ }\textbf {\bibinfo {volume} {96}},\
  \bibinfo {pages} {99} (\bibinfo {year} {1954})}\BibitemShut {NoStop}%
\bibitem [{\citenamefont {Yaresko}\ \emph {et~al.}(2009)\citenamefont
  {Yaresko}, \citenamefont {Liu}, \citenamefont {Antonov},\ and\ \citenamefont
  {Andersen}}]{pnictides-biq}%
  \BibitemOpen
  \bibfield  {author} {\bibinfo {author} {\bibfnamefont {A.~N.}\ \bibnamefont
  {Yaresko}}, \bibinfo {author} {\bibfnamefont {G.-Q.}\ \bibnamefont {Liu}},
  \bibinfo {author} {\bibfnamefont {V.~N.}\ \bibnamefont {Antonov}}, \ and\
  \bibinfo {author} {\bibfnamefont {O.~K.}\ \bibnamefont {Andersen}},\ }\href
  {\doibase 10.1103/PhysRevB.79.144421} {\bibfield  {journal} {\bibinfo
  {journal} {Phys. Rev. B}\ }\textbf {\bibinfo {volume} {79}},\ \bibinfo
  {pages} {144421} (\bibinfo {year} {2009})}\BibitemShut {NoStop}%
\bibitem [{Note1()}]{Note1}%
  \BibitemOpen
  \bibinfo {note} {$J_2^{\prime }$ manifests itself in the AFM state. It gives
  rise to the difference in the values of the effective parameters between
  parallel and antiparallel pairs of Fe spins.}\BibitemShut {Stop}%
\bibitem [{\citenamefont {Stanek}\ \emph {et~al.}(2011)\citenamefont {Stanek},
  \citenamefont {Sushkov},\ and\ \citenamefont {Uhrig}}]{mc-biquad-pnictides}%
  \BibitemOpen
  \bibfield  {author} {\bibinfo {author} {\bibfnamefont {D.}~\bibnamefont
  {Stanek}}, \bibinfo {author} {\bibfnamefont {O.~P.}\ \bibnamefont {Sushkov}},
  \ and\ \bibinfo {author} {\bibfnamefont {G.~S.}\ \bibnamefont {Uhrig}},\
  }\href {\doibase 10.1103/PhysRevB.84.064505} {\bibfield  {journal} {\bibinfo
  {journal} {Phys. Rev. B}\ }\textbf {\bibinfo {volume} {84}},\ \bibinfo
  {pages} {064505} (\bibinfo {year} {2011})}\BibitemShut {NoStop}%
\bibitem [{\citenamefont {Endoh}\ and\ \citenamefont
  {Ishikawa}(1971)}]{Endoh-3Q}%
  \BibitemOpen
  \bibfield  {author} {\bibinfo {author} {\bibfnamefont {Y.}~\bibnamefont
  {Endoh}}\ and\ \bibinfo {author} {\bibfnamefont {Y.}~\bibnamefont
  {Ishikawa}},\ }\href {\doibase 10.1143/JPSJ.30.1614} {\bibfield  {journal}
  {\bibinfo  {journal} {Journal of the Physical Society of Japan}\ }\textbf
  {\bibinfo {volume} {30}},\ \bibinfo {pages} {1614} (\bibinfo {year}
  {1971})}\BibitemShut {NoStop}%
\bibitem [{\citenamefont {Jo}(1983)}]{Jo-higher-order}%
  \BibitemOpen
  \bibfield  {author} {\bibinfo {author} {\bibfnamefont {T.}~\bibnamefont
  {Jo}},\ }\href {http://stacks.iop.org/0305-4608/13/i=10/a=006} {\bibfield
  {journal} {\bibinfo  {journal} {Journal of Physics F: Metal Physics}\
  }\textbf {\bibinfo {volume} {13}},\ \bibinfo {pages} {L211} (\bibinfo {year}
  {1983})}\BibitemShut {NoStop}%
\bibitem [{\citenamefont {Long}(1989)}]{Cu3Au-3Q-stability}%
  \BibitemOpen
  \bibfield  {author} {\bibinfo {author} {\bibfnamefont {M.~W.}\ \bibnamefont
  {Long}},\ }\href {http://stacks.iop.org/0953-8984/1/i=17/a=008} {\bibfield
  {journal} {\bibinfo  {journal} {Journal of Physics: Condensed Matter}\
  }\textbf {\bibinfo {volume} {1}},\ \bibinfo {pages} {2857} (\bibinfo {year}
  {1989})}\BibitemShut {NoStop}%
\bibitem [{Note2()}]{Note2}%
  \BibitemOpen
  \bibinfo {note} {We also verified that the inclusion of SOC does not modify
  the present phase diagram}\BibitemShut {NoStop}%
\bibitem [{\citenamefont {Kouvel}\ and\ \citenamefont
  {Kasper}(1963)}]{3Q-vs-AFM}%
  \BibitemOpen
  \bibfield  {author} {\bibinfo {author} {\bibfnamefont {J.}~\bibnamefont
  {Kouvel}}\ and\ \bibinfo {author} {\bibfnamefont {J.}~\bibnamefont
  {Kasper}},\ }\href {\doibase 10.1016/0022-3697(63)90148-3} {\bibfield
  {journal} {\bibinfo  {journal} {Journal of Physics and Chemistry of Solids}\
  }\textbf {\bibinfo {volume} {24}},\ \bibinfo {pages} {529 } (\bibinfo {year}
  {1963})}\BibitemShut {NoStop}%
\bibitem [{\citenamefont {Kawarazaki}\ \emph {et~al.}(1988)\citenamefont
  {Kawarazaki}, \citenamefont {Fujita}, \citenamefont {Yasuda}, \citenamefont
  {Sasaki}, \citenamefont {Mizusaki},\ and\ \citenamefont
  {Hirai}}]{3Q-how-to-measure}%
  \BibitemOpen
  \bibfield  {author} {\bibinfo {author} {\bibfnamefont {S.}~\bibnamefont
  {Kawarazaki}}, \bibinfo {author} {\bibfnamefont {K.}~\bibnamefont {Fujita}},
  \bibinfo {author} {\bibfnamefont {K.}~\bibnamefont {Yasuda}}, \bibinfo
  {author} {\bibfnamefont {Y.}~\bibnamefont {Sasaki}}, \bibinfo {author}
  {\bibfnamefont {T.}~\bibnamefont {Mizusaki}}, \ and\ \bibinfo {author}
  {\bibfnamefont {A.}~\bibnamefont {Hirai}},\ }\href {\doibase
  10.1103/PhysRevLett.61.471} {\bibfield  {journal} {\bibinfo  {journal} {Phys.
  Rev. Lett.}\ }\textbf {\bibinfo {volume} {61}},\ \bibinfo {pages} {471}
  (\bibinfo {year} {1988})}\BibitemShut {NoStop}%
\bibitem [{\citenamefont {Kennedy}\ and\ \citenamefont
  {Hicks}(1987)}]{moessbauer-on-3Q}%
  \BibitemOpen
  \bibfield  {author} {\bibinfo {author} {\bibfnamefont {S.~J.}\ \bibnamefont
  {Kennedy}}\ and\ \bibinfo {author} {\bibfnamefont {T.~J.}\ \bibnamefont
  {Hicks}},\ }\href {http://stacks.iop.org/0305-4608/17/i=7/a=015} {\bibfield
  {journal} {\bibinfo  {journal} {Journal of Physics F: Metal Physics}\
  }\textbf {\bibinfo {volume} {17}},\ \bibinfo {pages} {1599} (\bibinfo {year}
  {1987})}\BibitemShut {NoStop}%
\bibitem [{\citenamefont {Winterrose}\ \emph {et~al.}(2011)\citenamefont
  {Winterrose}, \citenamefont {Mauger}, \citenamefont {Halevy}, \citenamefont
  {Yue}, \citenamefont {Lucas}, \citenamefont {Mu\~noz}, \citenamefont {Tan},
  \citenamefont {Xiao}, \citenamefont {Chow}, \citenamefont {Sturhahn},
  \citenamefont {Toellner}, \citenamefont {Alp},\ and\ \citenamefont
  {Fultz}}]{winterrosePRB}%
  \BibitemOpen
  \bibfield  {author} {\bibinfo {author} {\bibfnamefont {M.~L.}\ \bibnamefont
  {Winterrose}}, \bibinfo {author} {\bibfnamefont {L.}~\bibnamefont {Mauger}},
  \bibinfo {author} {\bibfnamefont {I.}~\bibnamefont {Halevy}}, \bibinfo
  {author} {\bibfnamefont {A.~F.}\ \bibnamefont {Yue}}, \bibinfo {author}
  {\bibfnamefont {M.~S.}\ \bibnamefont {Lucas}}, \bibinfo {author}
  {\bibfnamefont {J.~A.}\ \bibnamefont {Mu\~noz}}, \bibinfo {author}
  {\bibfnamefont {H.}~\bibnamefont {Tan}}, \bibinfo {author} {\bibfnamefont
  {Y.}~\bibnamefont {Xiao}}, \bibinfo {author} {\bibfnamefont {P.}~\bibnamefont
  {Chow}}, \bibinfo {author} {\bibfnamefont {W.}~\bibnamefont {Sturhahn}},
  \bibinfo {author} {\bibfnamefont {T.~S.}\ \bibnamefont {Toellner}}, \bibinfo
  {author} {\bibfnamefont {E.~E.}\ \bibnamefont {Alp}}, \ and\ \bibinfo
  {author} {\bibfnamefont {B.}~\bibnamefont {Fultz}},\ }\href {\doibase
  10.1103/PhysRevB.83.134304} {\bibfield  {journal} {\bibinfo  {journal} {Phys.
  Rev. B}\ }\textbf {\bibinfo {volume} {83}},\ \bibinfo {pages} {134304}
  (\bibinfo {year} {2011})}\BibitemShut {NoStop}%
\end{thebibliography}
\end{document}